\documentclass[11pt]{article}

\usepackage[utf8]{inputenc}
\usepackage[iso-8859-1]{inputenx}
\usepackage{graphicx}
\usepackage[english]{babel}
\usepackage{amsmath,mathtools}
\usepackage{amssymb}
\usepackage{amsfonts}
\usepackage[margin=1in]{geometry}
\usepackage{hyperref}
\usepackage{float}
\usepackage{multirow}
\usepackage{latexsym}
\usepackage{caption}
\usepackage{babelbib}
\usepackage{latexsym}
\usepackage[T1]{fontenc}
\usepackage{tabularx}
\usepackage{dcolumn}
\usepackage{eurosym}
\usepackage{setspace}
\usepackage{color}
\usepackage[usenames,dvipsnames]{xcolor}
\usepackage{geometry}
\usepackage{multicol}
\usepackage{caption}
\usepackage{subfigure}
\usepackage{amsthm}
\usepackage{stmaryrd}
\usepackage{enumitem}
\usepackage{physics}
\usepackage{bbm}
\usepackage{bbding}
\usepackage{wasysym}
\usepackage{amsmath,accents}
\usepackage[dvipsnames]{xcolor}

\usepackage{mathrsfs}

\usepackage[T1]{fontenc}
\usepackage{tgbonum}
\usepackage{lmodern}

\usepackage{amssymb}
\usepackage{amssymb}
\usepackage{pifont}
\usepackage{float}

\restylefloat{table}

\numberwithin{equation}{section}

\newcommand{\im}[2]{{#1}_{\scriptscriptstyle{#2}}}

\newtheorem*{theorem*}{Theorem}

\newcommand{\e}[1]{\im{e}{#1}}

\let\OLDthebibliography\thebibliography
\renewcommand\thebibliography[1]{
  \OLDthebibliography{#1}
  \setlength{\parskip}{1.5pt}
  \setlength{\itemsep}{2pt plus 2ex}
}

\title{\textsf{\textbf{Geometrical Quantum Time in the $U(1)^3$ Model of Euclidean Quantum Gravity}}}
\author{
{\textrm{Sepideh Bakhoda}\thanks{s-bakhoda@bnu.edu.cn}\;, \textrm{Yongge Ma}\thanks{mayg@bnu.edu.cn}}
\\
\\
\textit{School of Physics and Astronomy, Key Laboratory of Multiscale Spin Physics} \\
\textit{(Ministry of Education), Beijing Normal University, Beijing 100875, China}}
\date{{\small\rm \today}}

\begin{document}

\maketitle
\sf{
\begin{abstract}
\rm{Loop Quantum Gravity faces challenges in constructing a well-defined Hamiltonian constraint and understanding the quantum notion of time. In this paper these issues are studied  by quantizing the $U(1)^3$ model, a simplified system exhibiting features similar to general relativity. By isolating a holonomy component within the Hamiltonian constraint, a discrete relative time evolution equation for quantum states is obtained. Then a Shr\"{o}dinger-like equation is derived in continuous limit. Thus the physical states solving this Shr\"{o}dinger-like equation can be written out. The emergence of the time parameter and its corresponding quantum operator are analyzed. It indicates the notion of a geometrical quantum time for quantum gravity.
}
\end{abstract}
}


\section{\textsf{Introduction}}
\rm{
Loop quantum gravity (LQG)  \cite{LQG} presents a compelling approach to attain the fusion of general relativity (GR) and quantum mechanics. The distinguished characteristic of LQG lies in its nonperturbative and background-independent framework whose kinematical Hilbert space comprises cylindrical functions over finite graphs embedded within the spatial manifold. While the Kinematics of LQG is well established and well understood, for over two decades, a central research focus has been on understanding the theory's quantum dynamics \cite{Thiemann:2023zjd}. In the canonical approach, Thiemann in his seminal papers known as QSD \cite{Thiemann:1996aw} proposed a suitable regularization procedure to obtain well-defined Hamiltonian constraint operators. 
However, the construction of the Hamiltonian constraint operator is not without ambiguity. In fact, these operators introduce new arcs and trivalent co-planar vertices to the graph corresponding to the cylindrical function upon which they act and this is where ambiguities arises in the form of dependence on the arbitrary choice of the new arcs.
Moreover, notwithstanding the absence of \textit{mathematical} anomalies in the original formulation, it should be noted that the precise structure functions found in the classical formulation are not necessarily recovered (contains \textit{physical} anomalies). This poses a complex challenge, as opposed to classical theory, where the quantum theory lacks a generator for spatial diffeomorphisms due to the discontinuity of 1-parameter unitary operator subgroups within the operators of finite spatial diffeomorphism group. Consequently, this issue holds great significance and warrants attention, as the Hamiltonian constraint operator is plagued by ambiguities that are expected to be resolved with the attainment of a faithful representation of the constraint algebra.
This issue has become a significant source of inspiration for the development of alternative methodologies (e.g., \cite{Thiemann:2005zg, Brown:1994py, Laddha:2014xsa, Thiemann:2020cuq}) aimed at resolving these ambiguities in the quantization process. To assess the effectiveness of these methods, it is highly beneficial to test them on other models that are simpler than general relativity but share its key features. In fact, by testing these methods on toy models, we can identify their shortcomings and seek ways to modify or extend them to apply to the full theory. One of these models, which we will focus on in this paper, is the U(1)$^3$ model \cite{Smolin:1992wj}. The feature that distinguishes this model from other toy models (e.g., \cite{Kuchar:1989bk, Husain:1990vz}) is that, in addition to sharing the main characteristics of background independence and general covariance with general relativity, the algebra of its constraints also forms an algebroid, rather than an algebra. In other words, the Poisson bracket between its constraints closes on structure functions, rather than structure constants, as is the case in the full theory.  The characteristics mentioned make the U(1)$^3$ model an interesting and special theory for further investigation using various quantization techniques. 

This model has been studied extensively from various aspects.
The electric shift method with a non-standard density weight for the Hamiltonian constraint was applied to this model in \cite{Tomlin:2012ejk}.
Its exact quantization while maintaining the standard density weight for the Hamiltonian constraint was achieved in \cite{Thiemann:2022all}. This approach is facilitated by the unique feature of the U(1)$^3$ model, where all constraints possess at most linear dependence in momentum. 
Building upon this feature, a classical-level solution for all constraints was employed to obtain the physical Hamiltonian in \cite{Bakhoda:2020ril}.
Moreover, the reduced phase space quantization of this model can be completed, demonstrating its consistency with the exact quantization \cite{Thiemann:2022all}. 
The model was further explored within the context of asymptotically flat spacetime in \cite{Bakhoda:2020eos}, with the findings subsequently utilized on reduced phase space approach. The covariant origin of the model was established in \cite{Bakhoda:2020fiy}, laying the groundwork for the development of a spin foam theory. Further steps for applying this concept to the U(1)$^3$ model were outlined in \cite{Thiemann:2022all}.
Finally, the effective dynamics of this theory was studied in \cite{Long:2021izw}.

While constructing the reduced phase space of GR explicitly proves challenging, the inclusion of a scalar field provides a suitable framework for defining the algebra of classical observables \cite{Brown:1994py}. 
A variety of matter fields can serve as ``clocks'' in this relational framework. In scenarios where the geometric constraints permit solving for certain momenta, these momenta can function as clocks without introducing additional matter. These are called geometric clocks. This strategy was employed for the $U(1)^3$ model in \cite{Thiemann:2022all, Bakhoda:2020ril}. Alternatively, a combination of geometric and matter degrees of freedom can define clocks, as explored in \cite{Giesel:2017roz} for cosmological observables. It is worth emphasizing that the choice of clock within the relational formalism is arbitrary. 

In contrast to the reduced phase space approach, in this paper we quantize the model using the Dirac quantization procedure. Subsequently, we isolate one holonomy component in the resulting expression for the Hamiltonian constraint. The objective is to recast the Hamiltonian constraint in the form of a discrete time evolution equation or a Shr\"{o}dinger-like equation in certain limit. This is achieved by identifying the conjugate variable to the isolated holonomy and interpreting it as a time parameter. The Shr\"{o}dinger-like equation introduces a time parameter, raising the fundamental question of whether a Hilbert space can be defined whose states satisfy the Shr\"{o}dinger-like equation. If the answer is affirmative, the solutions of this equation can serve as physical states and respect the notion of time evolution as well. Further questions arise: Given the Dirac quantization procedure and the quantization of all canonical variables, can the time parameter obtained in previous steps be quantized? Moreover, is the resulting time operator self-adjoint on the constructed Hilbert space? This paper presents an investigation into these issues.
\\
\\
Our paper is structured as follows:
In Section 2, following a brief review of the $U(1)^3$ model, we introduce several concepts that will be essential in subsequent sections. Section 3 is dedicated to the detailed derivation of all the elements required to construct the quantum operator corresponding to the Hamiltonian constraint. Subsequently, we introduce the Hamiltonian constraint operator. In Section 4, we recast the obtained operator in the form of a discrete time evolution equation as well as a Shr\"{o}dinger-like equation in a continuous limit, introducing a discrete time evolution. We then delve into the properties of time operator and introduce the Hilbert space whose states satisfying the Shr\"{o}dinger-like equation in continuous limit.
In the final section, we summarize and discuss our results.

\section{\textsf{Preliminaries}}
We begin by recalling that the (1+3)-dimensional Lorentzian Ashtekar-Barbero formalism of GR \cite{Ashtekar:1986yd} is constructed based on the Yang-Mills like variables $(\omega^i_a, E^a_i)$, where the configuration variable $\omega^i_a$ is the su(2)-valued connection field and the momentum variable $E^a_i$ is the densitized triad field, i.e. $E^a_i=\sqrt{\det(q)} e^a_i$, on a 3-dimensional spatial manifold $\Sigma$ whose metric is detoned by $q_{ab}$. Here, we employed the spatial indices $a, b \in \{1, 2, 3\}$ and the frame indices $i, j \in \{1,2,3\}$. The foundation of this phase space's symplectic structure rests upon the Poisson brackets defined for its canonical variables. Notably, the only non-vanishing bracket is $\{\omega_a^i(x), E^b_j(y) \}=\frac{\kappa \beta}{2} \delta^i_j \delta^b_a \delta(x, y)$, where $\kappa$ and $\beta$ represent the gravitational constant and Immirzi parameter \cite{Ashtekar:1986yd}, respectively. Then the action of GR in canonical Hamiltonian is obtained up to boundary terms as
\begin{equation}\label{Full action}
S[\omega, E, N, N, \Lambda]=\frac{1}{\kappa}\int \text{d}t \int_\Sigma \text{d}^3 x \left(E^a_i \dot{\omega}^i_a - \mathscr{H}[N] - \mathscr{H}_a[N^a] - \mathscr{G}_i[\Lambda^i] \right).
\end{equation}
So the dynamics is governed by the Hamiltonian which is the sum of the Gauss constraint $\mathscr{G}[\Lambda]$, the diffeomorphism constraint $\mathscr{H}_a[N^a]$, and the Hamiltonian constraint $\mathscr{H}[N]$, where $N, N^a $ and $\Lambda^i$ are Lagrange multipliers.
\\
\\
Three primary approaches exist to arrive at the U(1)$^3$ model:
\begin{itemize}
\item Direct Gauge Group Replacement: The internal gauge group SU(2) of GR formulated with Ashtekar-Barbero variables can be replaced with a direct product of three independent U(1) groups.
\item Zero Newton's Constant Approach: Alternatively, the Newtonian gravitational constant, $\kappa$, can be sent to zero at the action level. In fact, the canonical variables $(A^i_a, E^a_i)$ are defined through the Ashtekar-Barbero variables $(\omega^i_a, E^a_i)$ for the full theory by the scaling $A^i_a=\kappa^{-1} \omega^i_a$, which under the limit $\kappa \to 0$ effectively changes the local SU(2) frame rotation symmetry into the local U(1)$^3$ symmetry. Rewriting the action (\ref{Full action}) in terms of the scaled connection and then setting $\kappa \to 0$, it is easy to obtain at first order:
\begin{equation}\label{Hamiltonian Action}
S[A, E, N, N, \Lambda]=\int \text{d}t \int_\Sigma \text{d}^3 x \left(E^a_i \dot{A}^i_a - H[N] - H_a[N^a] - G_i[\Lambda^i] \right),
\end{equation}
where
\begin{equation}\label{Constraints in the model}
G[\Lambda]= \int_\Sigma \text{d}^3 x \; \Lambda^i \partial_a E^a_i, \;\;\; H_a[N^a]= \int_\Sigma \text{d}^3 x \; N^a F^i_{ab} E^b_i, 
\end{equation}
\begin{equation}\label{Classical Hamiltonian Constraint}
H[N]= \int_\Sigma \text{d}^3 x \; N \frac{\epsilon_{ijk}F^i_{ab} E^a_j E^b_k}{\sqrt{\det(q)}},
\end{equation}
with $F^i_{ab} = 2 \partial_{[a} A^i_{b]}$.
Thus the dynamics is governed by the Gauss, vector and scalar constraints respectively shown in (\ref{Constraints in the model}). 
It is noteworthy that the Gauss constraints presented in (\ref{Constraints in the model}) generate three independent U(1) gauge transformations on the connection fields, $A^1_a, A^2_a, A^3_a$. Consequently, the curvature $F^i_{ab}$ and the electric fields $E^a_i$ remain invariant under the action of the U(1)$^3$ group. This gauge invariance implies that the action (\ref{Hamiltonian Action}) describes a U(1)$^3$ theory, thereby justifying its designation.
\item Twisted Selfdual Action: The action of the twisted selfdual model is documented in \cite{Bakhoda:2020fiy} as
 \begin{equation}\label{action}
S= \int_\Sigma \text{d}^4 x \; \det(e) F^{IJ}_{A B} e^A_I e^B_J,
\end{equation}
with spacetime indices $A, B = 0, 1, 2, 3$ and internal indices $I, J = 0, \dots, 3$, where the determinant of the co-tetrad field $e^I_A$  is denoted by $\det(e)$, and $F^{IJ}$ is a twisted selfdual U(1)$^6$ curvature that is constrained by $ \frac{1}{2 \beta} \epsilon_{ijk} F^{jk}_{AB}= F^{0i}_{AB}=: F^i_{AB} = 2 \partial_{[A} A^i_{B]} $. As it has been shown in \cite{Bakhoda:2020fiy}, a Dirac constraint analysis can be conducted to arrive at the canonical theory (\ref{Hamiltonian Action}). Thus the action (\ref{action}) is the covariant origin of the U(1)$^3$ model. 
\end{itemize}
Therefore, the U(1)$^3$ model is characterized by a set of three independant U(1) connections denoted by $A^i_a$, along with associated electric fields denoted by $E^a_i$, where the electric fields are conceptualized to define a doubly densitized contravariant metric, similar to the interpretation in the context of gravity. The evolution of the model is governed by the constraints (\ref{Constraints in the model}). Furthermore, the symplectic structure is established through the fundamental Poisson bracket
\begin{equation}\label{poisson bracket}
\{A_a^i(x), E^b_j(y) \}= \frac{\beta}{2} \delta^i_j \delta^b_a \delta(x, y),
\end{equation}
where $\beta$ signifies the Immirzi parameter.
Note that the possibility existed to introduce three distinct Immirzi parameters, i.e. $\beta_1$, $\beta_2$, $\beta_3$, in the Poisson bracket (\ref{poisson bracket}). However, for the sake of simplicity and to maintain a parallel with the case of gravity, where a single Immirzi parameter governs the theory, we opted to set $\beta_1=\beta_2=\beta_3=\beta$.

The establishment of the kinematical Hilbert space for the U(1)$^3$ model draws a significant parallel to the well-established case of SU(2) gauge theory within LQG. 
Our point of departure lies in identifying the fundamental kinematical variables.  These variables are chosen to be U(1)$^3$-holonomies associated with edges and electric fluxes penetrating 2-surfaces within the spatial manifold (A detailed justification for the selection of holonomy-flux variables can be found in \cite{Bakhoda:2022rut}).
To mathematically express the concept of holonomy, let us consider a $C^k$  semianalytic embedded edge denoted by $e: [0,1]\to \Sigma$, where $k\gg1$. It is important to recall that due to the commutative nature of the U(1) group, all irreducible representations are one-dimensional.  Consequently, the $q$-representation of the U(1)-holonomy along edge $e$ can be expressed as $\pi_q(h_e (A))=\exp \left(i \kappa q \int_e A_a dx^a \right)$, where 
$q$ is an integer ($q\in \mathbb{Z}$) and $\kappa$ introduces a dimension of length multiplied by the inverse of mass and is necessary to ensure the exponent remains dimensionless. The product of three instances of these U(1)-holonomies serves as a representation of the U(1)$^3$-holonomy along the edge, i.e.
\begin{equation}\label{holonomy of U(1)3}
h_{e, \vec{q}}(A):= \prod_{i=1}^3  \pi_{q^i}(h_e (A^i)) =\exp \left(i \kappa   q_i \int_e A^i_a dx^a \right).
\end{equation}
Alongside holonomy, the other canonical variables are the U(1)$^3$ invariant flux variables for $E$, which are defined over an oriented 2-surface $S$ as
\begin{equation}\label{classical flux}
E_i(S)= \int_S \epsilon_{abc} E^a_i dx^b \wedge dx^c.
\end{equation}

The concept of holonomy, previously established for edges in (\ref{holonomy of U(1)3}), can be naturally extended to encompass entire graphs. Consider a closed, oriented graph denoted by $\gamma$. This graph is comprised of a collection of edges denoted by $\{e_I\}$, which exclusively meet at their endpoints, referred to as vertices. 
Within this framework, we can assign a set of labels $\{\vec{q}_I\}$ to each individual edge. Collectively, the graph and its edge labels are denoted by $c:=(\gamma, \{\vec{q}_I\})$. Then, the concept of \textit{graph holonomy}, denoted by $h_c(A)$, is defined as the product of holonomies along each individual edge, weighted by the corresponding label, i.e.
\begin{equation}
h_c(A):= \prod_I  h_{e_I, \vec{q}_I}(A) =\exp \left(i \kappa  \sum_I  q_I^i \int_{e_I} A^i_a dx^a \right)=: \exp \left(\int_{\Sigma}  \text{d}^3 x \; c^a_i A^i_a \right),
\end{equation}
where
\begin{equation}\label{def. c}
c^a_i (x; c) := \sum_I  i \kappa q_I^i \int \text{d} t_I \; \delta^3(x, e_I(t_I)) \dot{e}_I^a,
\end{equation}
in which $t_I \in [0,1]$ parametrizes the edge $e_I$ within the interval $[0,1]$, and $\dot{e}_I^a$  represents the derivative of the edge with respect to the parameter.
It can be verified (see \cite{Bakhoda:2022rut}) that graph holonomy, $h_c(A)$ exhibits invariance under U(1)$^3$ gauge transformations if and only if a specific condition holds. This condition applies to the edge labels $q^i_{I_v}$, for the set of all edges, $\{e_{I_v}\}$, that share a common vertex, $v$, in the graph.  In fact, for each $i$, it requires
\begin{equation}\label{gauge invariant condition}
\sum_I s(e_{I_v}) q^i_{I_v}=0,
\end{equation}
 where $s(e_{I_v})$ takes the value of $+1$ or $-1$ depending on whether the edge, $e_{I_v}$, is outgoing or incoming at vertex $v$, respectively. A graph holonomy, $h_c$, that satisfies this gauge invariance condition is referred to as a \textit{gauge-invariant charge network function} (CNF).

All Poisson brackets involving these holonomy and flux variables vanish except for $\{h_c, E_j (S)\}$. This specific bracket can be readily calculated as
\begin{equation}\label{holonomy flux algebra}
\{h_c, E_j (S)\} = i \kappa \beta\sum_{e_I \subset \gamma} \epsilon(e_I, S) q_j^I h_c.
\end{equation}
The underlying graph associated with the charge network $c$ is chosen in such a way ensuring that any isolated points where this graph intersects with surface S coincide with the graph's vertices.
The integer-valued function $\epsilon(e_I, S)$ is zero unless edge $e_I$ intersects surface $S$ transversally. In cases of a transverse intersection, the value is determined by the orientation: $\epsilon(e_I, S)=+1$ if edge $e_I$ is outgoing from and positioned above surface $S$ or incoming and positioned below $S$, and $\epsilon(e_I, S)=-1$ otherwise.
This refined definition of $\epsilon(e_I, S)$ ensures that it captures the appropriate orientation information for calculating the Poisson bracket.

We introduce the concept of an invariant charge network state, denoted by $|T_c\rangle$. This state represents a kinematic quantum state characterized by a graph holonomy $h_c$ which satisfies the condition (\ref{gauge invariant condition}).
The collection of all distinct charge network states forms the foundation of the U(1)$^3$ invariant kinematic Hilbert space, denoted by $\mathcal{H}_{\text{kin}}$. This space can be visualized as the span of a basis set $\{|T_c\rangle\}$.  Furthermore, an inner product is defined on $\mathcal{H}_{\text{kin}}$, given by  $\langle T_c | T_{c '} \rangle = \delta_{c, c'}$. The Kronecker delta function $\delta_{c, c'}$ ensures an inner product of zero for distinct states $(c\ne c')$ and one for identical states $(c = c')$.
It is important to acknowledge a potential issue with the labeling system for charge network states using $c:=(\gamma, \{\vec{q}_I\})$. This labeling suffers from a lack of uniqueness.  One can arbitrarily introduce additional vertices and edges that have no bearing on the underlying state itself, resulting in different labels for the same physical state.  To eliminate this redundancy and ensure consistency, we will adopt a standardized approach.  Each charge network state will be uniquely identified by its corresponding oriented graph with the minimum number of edges. 

Within this kinematical Hilbert space, a holonomy operator acts multiplicatively on a state, i.e.
\begin{equation}\label{holonomy operator}
\hat{h}_c |T_{c'}\rangle = |T_{c+c'}\rangle,
\end{equation}
where $c=(\gamma, \{\vec{q}\})$ and $c'=(\gamma', \{\vec{q}\;'\})$ and the resulting state, $|T_{c+c'}\rangle$, represents a new charge network state.  The associated graph of this new state is the minimal graph $\gamma''$ that encompasses both the original graphs $\gamma$ and $\gamma '$. Now, the graph $\gamma''$ can serve as a common underlying graph for both $|T_c \rangle$ and $|T_{c'}\rangle$. Next, we proceed by assigning the sum of the existing charge labels from states $|T_c \rangle$ and $|T_{c'}\rangle$ to the corresponding edges within the shared graph $\gamma''$. This process is performed in an ``edgewise'' manner, meaning each edge inherits a charge label of the original states.  The resulting graph with the new charge labels defines the newly formed charge network state $|T_c + T_{c'}\rangle$.

Considering the U(1)$^3$ model originated from the direct product of three distinct U(1)$_i$ models, we can extend the definition of holonomy operator corresponding for each U(1)$_i$ model in the kinematical Hilbert space of the U(1)$^3$ theory. As previously established, the holonomy operator for the U(1)$_i$ theory is defined by $h_{(\gamma, \{q^i_I\})} (A^i)=\exp \left(i \kappa \sum_I q^i_I \int_e A^i_a dx^a \right)$, where there is no summation over $i$. The corresponding holonomy in the U(1)$^3$ model is then given by $h_{{}^ic} (A)=\exp \left(i \kappa \sum_I\; {}^iq^j_I \int_e A^j_a dx^a \right)$, in which we have used the notations ${}^ic := (\gamma, \{{}^i\vec{q}_I\} )$ and ${}^i\vec{q}_I := (q^1_I \delta_1^i, q^2_I \delta_2^i, q^3_I \delta_3^i)$. ). This approach allows the transfer of holonomy associated with edge $e_I$ in the U(1)$_i$ model to the U(1)$^3$ theory, represented by the symbol $h^i_I$. Mathematically, this is expressed as
\begin{equation}
h^i_I = \exp \left(i \kappa \;{}^iq^j_I \int_e A^j_a dx^a \right).
\end{equation}
Consequently, its corresponding quantum operator acts on a quantum state as
 \begin{equation}\label{holonomy operator for Ai}
\hat{h}^i_I |T_c\rangle = |T_{{}^ic + c}\rangle,
\end{equation}
where the state $|T_{{}^ic + c}\rangle$ with ${}^ic = (e_I,  {}^i\vec{q}_I)$ is defined analogously to how $|T_{c+c'}\rangle$ was defined in (\ref{holonomy operator}). This concept will be crucial in the following section.

In contrast to holonomy, a flux operator acts on a state $T_c$ with $c=(\gamma, q^j_I)$ as a differential operator, i.e.
\begin{equation}\label{flux operator}
\hat{E}_j(S) |T_c\rangle = \kappa \beta \hbar \sum_{v\in S} \sum_{\substack{I \\e_I \cap \{v\}\ne\emptyset }} \epsilon(e_I, S) q^j_I |T_c\rangle.
\end{equation}
One can demonstrate that the defined operators (\ref{holonomy operator}) and (\ref{flux operator}) establish a representation of the holonomy-flux algebra (\ref{holonomy flux algebra}) on $\mathcal{H}_{\text{kin}}$.

Therefore, similar to the framework of LQG \cite{LQG}, the fundamental elements for describing the quantum geometry are the holonomy and flux operators.
Furthermore, again similar to the kinematic Hilbert space of LQG, here $\mathcal{H}_{\text{kin}}$ allows for a natural representation of spatial diffeomorphisms. This is achieved by considering diffeomorphic deformations of the underlying graphs associated with the charge network states, while preserving the assigned charge labels.
The inner product defined in $\mathcal{H}_{\text{kin}}$ enables a group averaging procedure that ultimately yields a diffeomorphism-invariant Hilbert space, denoted by $\mathcal{H}_{\text{kin}}^{\text{diff}}$ which is a subspace of the dual space of the cylindrical functions over graphs. However, quantizing the remaining scalar constraints is a significantly more complex process.  This process necessitates a specialized regularization technique to facilitate the use of holonomy and flux variables. The scalar constraint operator will be constructed in $\mathcal{H}_{\text{kin}}$.
\\
\\
We consider an open set $\mathcal{U}$ within $\Sigma$ equipped with a coordinate system $\{x\}$. This neighborhood serves as a domain for further analysis. For any point $p$ residing in $\mathcal{U}$, its coordinates are denoted by $x(p)=\{x^1, x^2, x^3\}$. A coordinate ball centered at $p$ with radius $\varepsilon$ is denoted by $B_\varepsilon(p)$. This ball is a subset of the corresponding neighborhood $\mathcal{U}$. The volume of this ball is then expressed as $V_{(p; \varepsilon)}:= V({B_\varepsilon(p)})$, i.e.
\begin{equation}\label{volume of the Ball}
V_{(p; \varepsilon)}:= V({B_\varepsilon(p)}) = \int_{B_\varepsilon(p)} \text{d}x^3 \; \sqrt{q} = \int_{B_\varepsilon(p)} \text{d}x^3 \; \sqrt{\left|\det(E)\right|} =\int_{B_\varepsilon(p)} \text{d}x^3 \; \sqrt{\left|\frac{1}{3!} \epsilon_{abc} \epsilon^{ijk} E^a_i E^b_j E^c_k \right|},
\end{equation}
where $q$ denotes the determinant of the metric tensor.
We can also partition the neighborhood $\mathcal{U}$ using a collection $\mathscr{C}$ of closed cubes $\mathcal{C}$. These cubes are aligned parallel to the coordinate planes. If a cube $\mathcal{C}$ falls partially outside of region $B_\varepsilon(p)$, only its intersection with $B_\varepsilon(p)$ is considered for the purposes of this analysis. Notably, any two distinct cubes in this collection can only share points along their boundaries (they do not overlap in their interiors).
Within each individual cube $\mathcal{C}$ belonging to the collection $\mathscr{C}$, we introduce an ordered triplet $s = (S_1, S_2, S_3)$. This triplet consists of three oriented, two-dimensional surfaces (without boundaries). Each surface is defined by the equation $x^a =$ constant, where $a$ can take values $1, 2,$ and $3$. These surfaces intersect within the interior of the corresponding cube $\mathcal{C}$. The orientation of each surface is determined by the orientation of the corresponding coordinate axes. Then we can approximate $\det(E)$ at any internal point of the cell-cube $\mathcal{C}$ using the fluxes defined in (\ref{classical flux}) by
\begin{equation}\label{det E in terms of fluxes}
\det(E)_{\mathcal{C}}=\frac{1}{3!} \epsilon_{abc} \epsilon_{ijk} E^i(S^a) E^j(S^b) E^k(S^c).
\end{equation}
This approximation of the quantity $\det(E)$ naturally leads to an approximate expression for the volume of the ball (\ref{volume of the Ball}) associated with the partition $\mathscr{C}$. This expression is defined as the sum over all cells $\mathcal{C}$ in the partition of the absolute values of the smeared electric fields $\det(E)_{\mathcal{C}}$ within each cell $\mathcal{C}$, i.e. $V^{\mathscr{C}}_{(x, \varepsilon)} = \sum_{\mathcal{C}\in \mathscr{C}} \sqrt{\left| \det(E)_{\mathcal{C}} \right|}$.
In essence, this demonstrates that within classical theory, the phase space function $V_{(x, \varepsilon)}$ can be expressed in terms of fluxes. Since we already possess the quantum operators corresponding to the fluxes in (\ref{flux operator}), we can construct the regulated quantum operators of (\ref{det E in terms of fluxes}) as
\begin{equation}
\widehat{\det(E)}_{\mathcal{C}} |T_c\rangle = (\kappa \beta \hbar)^3 \sum_{\substack{I, J, K \\ e_I \cap e_J \cap e_K = \{v\}}} \frac{1}{6}\;\epsilon_{abc} \;\epsilon(e_I, S^a) \epsilon(e_J, S^b) \epsilon(e_K, S^c)  \;q_{IJK} \; |T_c\rangle,
\end{equation}
where 
\begin{equation}
q_{IJK}:= \epsilon_{ijk} q^i_I q^j_J q^k_K.
\end{equation}
However, removing these regulators is an intricate process. We refer the reader to \cite{Ashtekar:1997fb} for the details and subtleties regarding to the definition of permissible regulators. Here, we assume that the regulator is permissible in the sense of \cite{Ashtekar:1997fb}, and we suffice to mention that the action of the operator $\widehat{\det(E)}_{\mathcal{C}}$ on $|T_c\rangle$ hinges solely on the properties of the three 2-surfaces, $S^a$, only at vertex $v$. Consequently, refining the partition and shrinking cell $\mathcal{C}$ to solely encompass vertex $v$ does not alter the operator, i.e.
\begin{equation}
\widehat{\det(E)}(v) |T_c\rangle = \lim_{\mathcal{C}\to \{v \}} \widehat{\det(E)}_{\mathcal{C}} |T_c\rangle = \frac{(\kappa \beta \hbar)^3}{6}  \sum_{\substack{I, J, K \\ e_I \cap e_J \cap e_K = \{v\}}} \epsilon_{abc} \;\epsilon(e_I, S^a) \epsilon(e_J, S^b) \epsilon(e_K, S^c)  \;q_{IJK} \; |T_c\rangle,
\end{equation}
where $\epsilon(e_I, e_J, e_K):= \epsilon_{abc} \;\epsilon(e_I, S^a) \epsilon(e_J, S^b) \epsilon(e_K, S^c)$. However, this operator retains a dependence on the specific regulators employed, causing the resulting volume operator to lack diffeomorphism covariance.
To address this issue, a procedure for averaging over the relevant background structures was introduced in \cite{Ashtekar:1997fb}. Notably, the requirement for the final volume operator $\hat{V}_{(p, \varepsilon)}$ to be both well-defined and diffeomorphism covariant is so stringent that it dictates the form of the averaged operator, except for a single multiplicative constant denoted by $\mathfrak{C}_{\text{AL}}$. Therefore,
\begin{equation}\label{volume operator}
\hat{V}_{(p, \varepsilon)} |T_c\rangle = \mathfrak{C}_{\text{AL}} \left(\kappa \beta \hbar \right)^{\frac{3}{2}}\sum_{v\in B_\varepsilon(p) \cap V(\gamma)} \sqrt{\left| \mathcal{Q}_v \right|} \; |T_c\rangle,
\end{equation}
where
\begin{equation}\label{Q of v}
\mathcal{Q}_v := \sum_{\substack{I, J, K \\ e_I \cap e_J \cap e_K = \{v\}}} \frac{1}{6}\;\epsilon(e_I, e_J, e_K)  q_{IJK}.
\end{equation}
The term $\epsilon(e_I, e_J, e_K)$ represents an orientation function: if the three tangent directions associated with the triplet $(e_I, e_J, e_K)$ lie on the same plane at vertex $v$, the function vanishes. Conversely, the function takes a value of $+1$ or $-1$ if the tangent directions are linearly independent and form a positively or negatively oriented basis, respectively.

A noteworthy property of the volume operator (\ref{volume operator}) is that it excludes contributions from vertices $v$ of $\gamma$ whose edge tangents span, at most, a single plane. This characteristic arises from the term $\mathcal{Q}_v$ defined in (\ref{Q of v}), which demonstrably vanishes when $v$ represents a planar vertex.


\section{\textsf{Hamiltonian constraint}}
The process of obtaining the quantum operator corresponding to a Hamiltonian constraint involves a structured approach. This entails identifying the quantum operators for each constituent component within the constraint. These components typically include the holonomy, electric field, and the inverse determinant of the metric. Subsequently, these classical terms are replaced with their corresponding quantum operators within the Hamiltonian constraint. Finally, a thorough analysis is conducted to identify any potential ambiguities that may arise due to the introduction of quantum operators. This rigorous approach ensures a well-defined quantum representation of the classical constraint. 

Having established the quantum operator corresponding to the holonomy in (\ref{holonomy operator}), we proceed to derive the quantum operators for the electric field and the inverse square root of the metric determinant. As expected, the holonomy operator acts on a CNF by multiplication and electric field by differentiation, $\hat{E}^a_i \sim -i \hbar \frac{\delta}{\delta A^i_a}$. Thus,
\begin{equation}
\hat{E}_i^a(x) |T_c\rangle = -i \hbar \frac{\delta}{\delta A^i_a (x)} \exp \left( \int \text{d}^3y \, c^b_j(y) A_b^j (y) \right) = -i\hbar c^a_i(x) |T_c\rangle.
\end{equation}
The action of the electric field operator on a CNF is distributional, as detailed in the definition of $c_i^a$ in (\ref{def. c}). To address this, a regularization procedure is implemented.

We define a small neighborhood, denoted by $\mathcal{U}$, around each vertex $v$ belonging to the vertex set $V$ of the underlying graph.
Within each neighborhood $\mathcal{U}$, a coordinate chart ${x}$ is introduced. This chart specifies a local coordinate system where the subscript $v$ indicates that the origin is located at vertex $v$. 
We regularize the Dirac delta function using normalized characteristic functions of the coordinate balls with radius $\varepsilon$, i.e.
$\delta_\varepsilon(x, y) = \frac{3}{4 \pi \varepsilon^3} \chi_{B_\varepsilon(x)}(y)$, using which the regularized electric field operator can be calculated as
\begin{align}\label{Flux operator action}
\hat{E}^a_{i,\varepsilon}(v) |T_c\rangle &= -i\hbar c^a_{i,\varepsilon}(v) |T_c\rangle 
= \frac{-3i\hbar}{4\pi \varepsilon^3} \sum_I i q^i_I \int dt_I \; \chi_{B_\varepsilon(v)} (e_I(t_I)) \dot{e}_I^a(t_I) |T_c\rangle \nonumber \\
&= \frac{3\hbar}{4\pi \varepsilon^2} \sum_{e_I \cap \{v\}\ne \emptyset} q^i_I \hat{e}^a_I |T_c\rangle + O(\varepsilon^{-1})
\end{align}
where $\hat{e}^a_I$ is the unit coordinate tangent vector to $e_I$ at $v$. 

To find the quantum operator corresponding to the regularized version of $q^{-1/2}$, we will follow an approach that mirrors the principles employed in Thiemann's trick. 
We begin with the derivation of some classical identities which will be leveraged to facilitate the transition into the quantum realm.

The initial step involves derivation of the volume operator in the U(1)$^3$ model. 
Classically, the volume $V(R)$ of a region $R$ contained within $\Sigma$ is defined as $V(R)=\int_R d^3x \sqrt{|\det(E)|} $.
 We define $B_\varepsilon(x)\subset \Sigma$ as a coordinate ball centered at $x$ with radius $\varepsilon$ and $V_{(x; \varepsilon)}:=V(B_\varepsilon(x)) $. Then, one has
\begin{align}\label{PB of A and V}
\{A^k_c (x), V_{(x; \varepsilon)}^{\frac{1}{3}}\} &= \frac{\beta}{12} \sqrt{|q|} \;V_{(x; \varepsilon)}^{-\frac{2}{3}} E^k_c (x)\nonumber\\
& \sim \frac{\beta}{12} \sqrt{|q|} \;\left(q^{-\frac{1}{3}}(\frac{4\pi}{3}\varepsilon^3)^{-\frac{2}{3}}\right) E^k_c (x)\nonumber\\
& \sim  \frac{\beta}{\varepsilon^2} \left(\frac{\pi}{18 \sqrt{3}}\right)^{-\frac{2}{3}} q (x)^{\frac{1}{6}} E^k_c,
\end{align}
where we have used the approximation $V_{(x; \varepsilon)} \sim q^{\frac{1}{2}}(\frac{4\pi}{3}\varepsilon^3)$. The form of (\ref{PB of A and V}) suggests that in order to derive a suitable classical expression for $q (x)^{\frac{1}{2}}$ we need to multiply three terms of this form. It can readily be verified that
\begin{align}
\epsilon^{abc}\epsilon_{ijk}\{A^i_a (x), V_{(x; \varepsilon)}^{\frac{1}{3}}\}\{A^j_b (x), V_{(x; \varepsilon)}^{\frac{1}{3}}\}\{A^k_c (x), V_{(x; \varepsilon)}^{\frac{1}{3}}\} &=\frac{\beta^3}{\varepsilon^6} \left(\frac{\pi}{18 \sqrt{3}}\right)^{-2} q ^{\frac{1}{2}} \epsilon^{abc}\epsilon_{ijk} E^i_a  E^j_b E^k_c \nonumber\\
 &= \frac{\beta^3}{\varepsilon^6} \left(\frac{\pi}{18 \sqrt{3}}\right)^{-2} q(x)^{-\frac{1}{2}},
\end{align}
which leads to
\begin{equation}\label{classical q neg before hol}
q(x)^{-\frac{1}{2}} = \frac{\varepsilon^6}{\beta^3} \left(\frac{\pi}{18 \sqrt{3}}\right)^{2}  \epsilon^{abc}\epsilon_{ijk}\{A^i_a (x), V_{(x; \varepsilon)}^{\frac{1}{3}}\}\{A^j_b (x), V_{(x; \varepsilon)}^{\frac{1}{3}}\}\{A^k_c (x), V_{(x; \varepsilon)}^{\frac{1}{3}}\}.
\end{equation}
At this stage, it appears that by utilizing the rule $\{\cdot , \cdot\} \to (i\hbar)^{-1} [\cdot , \cdot]$ and substituting the remaining components with their quantum counterparts, one can obtain a quantum operator for $q(x)^{-\frac{1}{2}}$. However, the challenge lies in the absence of a well-defined operator for the connection $A^i_a$. Consequently, one must replace it with an approximation of the holonomy and subsequently employ the operator for the holonomy defined in (\ref{holonomy operator}) or (\ref{holonomy operator for Ai}) in the final expression. One can easily verify that 
\begin{align}
\{h_I^i, V_{(x; \varepsilon)}^{\frac{1}{3}}\} &= \int \text{d}^3 y \; \frac{\delta h_I^i}{\delta A^j_a (y)} \{A^j_a (y), V_{(x; \varepsilon)}^{\frac{1}{3}}\} 
\sim h_I^i \left(i \kappa \varepsilon q^i_I \hat{e}^a_I \{A^i_a (x),  V_{(x; \varepsilon)}^{\frac{1}{3}}\} \right),
\end{align}
where there is no summation over $i$ and $I$. Assuming that the integer number $q^i_I$ is not zero, one can arrive at
\begin{equation}\label{e times PB A and V}
\hat{e}^a_I \{A^i_a(x), V_{(x; \varepsilon)}^{\frac{1}{3}} \} \sim \frac{1}{i\kappa \varepsilon q^i_I} (h_I^i)^{-1} \{h_I^i,  V_{(x; \varepsilon)}^{\frac{1}{3}} \}.
\end{equation}
In pursuit of quantizing the expression (\ref{classical q neg before hol}) as an operator acting on the state $|T_c\rangle \in \mathcal{H}_{\text{kin}}$ with $c=(\gamma, \{\vec{q}_I\})$, We consider a specific scenario involving a triplet of edges, denoted by $e_I$ (where $I = 1, 2, 3$), emanating from a vertex $v$ within the underlying graph $\gamma$. Each edge has a coordinate length of $\varepsilon$ and must satisfy the condition that their tangents at $v$ form a linearly independent set.  This ensures a non-degenerate configuration and the invertibility of the matrix $\hat{e}^a_I$, i.e. $\hat{e}:= \det(\hat{e}^a_I)\ne 0$, such that one can isolate $\{A^i_a(x), V_{(x; \varepsilon)}^{\frac{1}{3}} \}$ in (\ref{e times PB A and V}) using the inverse of $\hat{e}^a_I$
\begin{equation}\label{PB A and V}
 \{A^i_a(x), V_{(x; \varepsilon)}^{\frac{1}{3}} \} \sim \frac{1}{i\kappa \varepsilon q^i_I}  \hat{e}^I_a \; (h_I^i)^{-1} \{h_I^i,  V_{(x; \varepsilon)}^{\frac{1}{3}} \}.
\end{equation}
Under these conditions and by substituting (\ref{PB A and V}) into (\ref{classical q neg before hol}), we arrive at the following identity
\begin{align}
q(v)^{-\frac{1}{2}} &=
\frac{i \varepsilon^3}{\kappa^3 \beta^3} \left(\frac{\pi}{18 \sqrt{3}}\right)^{2} \frac{1}{q^i_I q^j_J q^k_K} \epsilon^{abc}\epsilon_{ijk} \hat{e}^I_a \hat{e}^J_b \hat{e}^K_c \nonumber\\
&\hspace{1cm}  \cross \; (h_I^i)^{-1} \{h_I^i,  V_{(x; \varepsilon)}^{\frac{1}{3}} \}
  \; (h_J^j)^{-1} \{h_J (A^j),  V_{(x; \varepsilon)}^{\frac{1}{3}} \} 
  \; (h_K^k)^{-1} \{h_K (A^k),  V_{(x; \varepsilon)}^{\frac{1}{3}} \} \nonumber\\
  &= 
 \frac{i \varepsilon^3}{\kappa^3 \beta^3}\frac{\alpha}{q^i_I q^j_J q^k_K} \epsilon_{ijk} \epsilon^{IJK}  \; (h_I^i)^{-1} \{h_I^i,  V_{(x; \varepsilon)}^{\frac{1}{3}} \}
  \; (h_J^j)^{-1} \{h_J^j,  V_{(x; \varepsilon)}^{\frac{1}{3}} \} 
  \; (h_K^k)^{-1} \{h_K^k,  V_{(x; \varepsilon)}^{\frac{1}{3}} \},
\end{align}
where $\alpha := \left(\pi/18 \sqrt{3}\right)^{2} \hat{e}^{-1}$.
We now proceed to construct an $\varepsilon$-regularized operator acting on $\mathcal{H}_{\text{kin}}$. This is accomplished by promoting all quantities to their quantum counterparts and by using the rule $\{\cdot , \cdot\} \to (i\hbar)^{-1} [\cdot , \cdot]$. 
\begin{align}\label{q operator}
\hat{q}^{-\frac{1}{2}} &=
 \frac{- \varepsilon^3}{(\kappa \beta \hbar)^3}\frac{\alpha}{q^i_I q^j_J q^k_K} \epsilon_{ijk} \epsilon^{IJK} 
\; (\hat{h}_I^i)^{-1} [\hat{h}_I^i,  \hat{V}_{(x; \varepsilon)}^{\frac{1}{3}}]
  \; (\hat{h}_J^j)^{-1} [\hat{h}_J^j,  \hat{V}_{(x; \varepsilon)}^{\frac{1}{3}}]
  \; (\hat{h}_K^k)^{-1} [\hat{h}_K^k,  \hat{V}_{(x; \varepsilon)}^{\frac{1}{3}}].
\end{align}

To understand how this operator acts on the quantum state $|T_c\rangle$, we need to determine the action of the term $(\hat{h}_I^i)^{-1} [\hat{h}_I^i,  \hat{V}_{(x; \varepsilon)}^{\frac{1}{3}}]$ on this state. Here, $e_I$ represents an edge within the underlying graph of $|T_c\rangle$, and $\varepsilon$ is assumed to be sufficiently small such that a specific ball $B_\varepsilon (p)$ contains only one vertex $v$ of the graph $\gamma$. Under these assumptions, we have
\begin{equation}
\hat{V}_v |T_c\rangle:= V_{(v; \varepsilon)}\big|_{B_\varepsilon (v)\cap V(\gamma) = \{v\}} |T_c\rangle= \mathfrak{C}_{\text{AL}} \left(\kappa \beta \hbar \right)^{\frac{3}{2}} \sqrt{\left| \mathcal{Q}_v \right|} \; |T_c\rangle.
\end{equation}
Furthermore, we denote the eigenvalue of $\hat{V}_v$ acting on  $|T_{{}^ic+c}\rangle= h_I^i |T_c\rangle$ by 
\begin{equation}
\hat{V}_v|T_{{}^ic+c}\rangle = \mathfrak{C}_{\text{AL}} \left(\kappa \beta \hbar \right)^{\frac{3}{2}} \sqrt{\left| (\mathcal{Q}_v)^i_I \right|} |T_{{}^ic+c}\rangle.
\end{equation}
Utilizing these relationships, we can express the action of the aforementioned term on the state $|T_c\rangle$ as 
\begin{align}\label{part of q on state}
(\hat{h}_I^i)^{-1} [\hat{h}_I^i,  \hat{V}^{\frac{1}{3}}_{v}]|T_c\rangle &= \hat{V}^{\frac{1}{3}}_{v} |T_c\rangle - (\hat{h}_I^i)^{-1} \hat{V}^{\frac{1}{3}}_{v} \hat{h}_I^i |T_c\rangle \nonumber\\
&= \mathfrak{C}_{\text{AL}}^{\frac{1}{3}} \left(\kappa \beta \hbar \right)^{\frac{1}{2}} |\mathcal{Q}_v|^{\frac{1}{6}} |T_c\rangle - \mathfrak{C}_{\text{AL}}^{\frac{1}{3}} \left(\kappa \beta \hbar \right)^{\frac{1}{2}} \left| (\mathcal{Q}_v)^i_I \right|^{\frac{1}{6}} (\hat{h}_I^i)^{-1} |T_{{}^ic+c}\rangle \nonumber\\ 
&=\mathfrak{C}_{\text{AL}}^{\frac{1}{3}} \left(\kappa \beta \hbar \right)^{\frac{1}{2}} \left( |\mathcal{Q}_v|^{\frac{1}{6}} - \left|(\mathcal{Q}_v)^i_I \right|^{\frac{1}{6}} \right) |T_c\rangle.
\end{align}
Finally, by plugging this expression into the equation for the operator (\ref{q operator}), we obtain the following result for the action of the $\varepsilon$-regularized operator of $q^{-1/2}$ on the state $|T_c\rangle$ as
\begin{align}\label{operator of q -1/2}
\widehat{q^{-1/2}_\varepsilon(v)} |T_c\rangle = \frac{ \varepsilon^3}{\left(\kappa \beta \hbar \right)^{\frac{3}{2}}} \Upsilon (v) \; |T_c\rangle,
\end{align}
where
\begin{equation}
\Upsilon (v) :=- \frac{\alpha \mathfrak{C}_{\text{AL}}  \epsilon_{ijk} \epsilon^{IJK} }{q^i_I q^j_J q^k_K} 
\left( |\mathcal{Q}_v|^{\frac{1}{6}} - \left|(\mathcal{Q}_v)^i_I \right|^{\frac{1}{6}} \right)
\left( |\mathcal{Q}_v|^{\frac{1}{6}} - \left|(\mathcal{Q}_v)^j_J \right|^{\frac{1}{6}} \right)
\left( |\mathcal{Q}_v|^{\frac{1}{6}} - \left|(\mathcal{Q}_v)^k_K \right|^{\frac{1}{6}} \right).
\end{equation}

To promote the Hamiltonian constraint to an operator, it is necessary to express the curvature in terms of holonomies. This arises from the presence of the connection, through its curvature tensor $F_{ab}$, within the Hamiltonian constraint itself. As holonomy operators are the fundamental building blocks in this context, expressing curvature in terms of these operators becomes essential.

In the classical theory, components of the curvature tensor $F_i$ can be derived by computing holonomies around appropriately chosen loops. Subsequently, these holonomies are divided by the enclosed surface area and the limit of vanishing area is taken. For instance, to obtain $F_{xy}$, one considers a loop centered at point $x$ with infinitesimal coordinate area $\varepsilon^2$ lying within the $xy$-plane. Expanding the connection $A_a(\alpha(t))$ around $x_0$ yields $A_a(\alpha(t)) = A_a(x_0) + (\alpha(t) - x_0)^b \partial_b A_a(x_0) + O(\varepsilon^2)$. From this expansion, the holonomy around the loop can be computed as 
\begin{align}
h^i_{\alpha_{IJ}} &:= h_{\alpha_{IJ}}(A^i) = \exp \left(i \kappa  \int_0^1 \text{d}t \; A^i_a(\alpha(t)) \dot{\alpha}^a(t) \right) \nonumber\\
&=
1+ i\kappa  \int_0^1 \text{d}t \; A^i_a(\alpha(t)) \dot{\alpha}^a(t) \nonumber\\
& \quad \quad + (i\kappa)^2 \int_0^1 \text{d}t \int_0^t \text{d}t' \; A^i_a(\alpha(t)) \dot{\alpha}^a(t) A^j_b(\alpha(t')) \dot{\alpha}^b(t') + O(\varepsilon^3) \nonumber \\
&= 
1+ i\kappa \left(A^i_a (x_0) \int_0^1 \text{d}t\; \dot{\alpha} (t) + \partial_b A^i_a (x_0) \int_0^1 \text{d}t \;\dot{\alpha}^a (t) (\alpha^b(t)-x_0^b)\right) + O(\varepsilon^3)\nonumber\\
&=
1- i\kappa \varepsilon^2  F^i_{ab} (x_0)  + O(\varepsilon^3).
\end{align}
 Here, we have utilized the conditions $\alpha(0) = \alpha(1) = x_0$ and the fact that $\int_0^1 dt \dot{\alpha}^x(t)(\alpha^y(t) - x^y)$ represents the enclosed area of the loop. The appearance of curvature is intrinsically linked to the antisymmetric property of $\int_0^1 dt \dot{\alpha}^a(t)(\alpha^b(t) - x^b)$ with respect to indices $a$ and $b$. Generalizing to a loop $\alpha$ lying in the $x^a$-$x^b$ plane, we obtain
 \begin{equation}
\hat{F}^i_{ab, \varepsilon} (v) = \frac{1}{\varepsilon^2 } \frac{1}{i\kappa} \left(1- \hat{h}^i_{\alpha_{ab}} \right) +  O(\varepsilon).
\end{equation}

Having determined the corresponding operators for all constituents of the Hamiltonian constraint, we proceed to substitute these into (\ref{Classical Hamiltonian Constraint}) to construct its quantum counterpart. It is essential to preserve the operator ordering as it appears in (\ref{Classical Hamiltonian Constraint}), a decision that will be justified subsequently.

To establish a suitable framework, we introduce a triangulation $T(\varepsilon)$ of the spatial manifold $\Sigma$, parameterized by $\varepsilon$. This parameter regulates the granularity of the triangulation, with the limit $\varepsilon \rightarrow 0$ representing a continuum limit. Within each tetrahedron $\Delta \in T(\varepsilon)$, we designate a specific vertex $v(\Delta)$. Emanating from $v(\Delta)$, we define three edges, denoted by $s_a(\Delta)$ for $a = x, y, z$. These edges originate at $v(\Delta)$, and their tangents collectively define the tangent space at $v(\Delta)$. Additionally, we introduce arcs $s_{ab}(\Delta)$ connecting the endpoints of $s_a(\Delta)$ and $s_b(\Delta)$.
Utilizing these elements, we construct loops $\alpha_{ab}(\Delta) = s_a(\Delta) \circ s_{ab}(\Delta) \circ s_b(\Delta)^{-1}$ for each combination of $a$, $b$, and $\Delta$. The ensemble comprising $T(\varepsilon)$, $v(\Delta)$, $s_a(\Delta)$, and $\alpha_{ab}(\Delta)$ is termed a regulator.

Recognizing that the volume operator in the quantum counterpart of (\ref{Classical Hamiltonian Constraint}) exclusively acts upon the vertices of the underlying graph $\gamma$, a naive discretization might yield a trivial continuum limit. To circumvent this issue, we employ a triangulation adapted to the graph $\gamma$. This entails that each tetrahedron $\Delta$ contains at most one vertex of $\gamma$, and the outgoing edges $s_a(\Delta)$ align with edges of $\gamma$ incident at this vertex. Consequently, the loops $\alpha_{ab}(\Delta)$ are formed by segments of these graph edges and the connecting arcs $s_{ab}(\Delta)$.

Given these considerations, the operator corresponding to the regulated Hamiltonian constraint can be calculated as follows. Note that the operators $\hat{E}^a_{i, \varepsilon} (v)$ and $\widehat{q^{-1/2}_\varepsilon}(v)$ are diagonal according to (\ref{flux operator}) and (\ref{operator of q -1/2}) respectively, so we use simply their eigenvalues in the following.
\begin{align}\label{H[N] operator}
\widehat{H[N]}_\varepsilon |T_c\rangle &= \sum_{v\in V(\gamma)} \varepsilon^3 \; N(v) \epsilon_{ijk}\hat{F}^i_{ab, \varepsilon} \hat{E}^a_{j, \varepsilon}(v) \hat{E}^b_{k, \varepsilon} (v) \widehat{q^{-1/2}_\varepsilon}(v) |T_c\rangle \nonumber\\
&=
 \sum_{v\in V(\gamma)} \varepsilon^3 \; N(v) \epsilon_{ijk} \left(\frac{3\hbar}{4\pi \varepsilon^2} \sum_{e_I \cap \{v\}\ne \emptyset} q^j_I \hat{e}^a_I \right) \left(\frac{3\hbar}{4\pi \varepsilon^2} \sum_{e_J \cap \{v\}\ne \emptyset} q^k_J \hat{e}^b_J \right) \left(\frac{ \varepsilon^3}{\left(\kappa \beta \hbar \right)^{\frac{3}{2}}} \Upsilon \right) \hat{F}^i_{ab, \varepsilon} |T_c\rangle \nonumber\\
 &=
 \sum_{v\in V(\gamma)}  \frac{-9i \hbar^{\frac{1}{2}}}{16\pi^2 \kappa^{\frac{5}{2}} \beta^{\frac{3}{2}}}\; N(v) \epsilon_{ijk} \left(\sum_{e_I \cap \{v\}\ne \emptyset} q^j_I \hat{e}^a_I \right) \left(\sum_{e_J \cap \{v\}\ne \emptyset} q^k_J \hat{e}^b_J \right)  \Upsilon \;  \left(1- \hat{h}^i_{\alpha_{ab}} \right) |T_c\rangle \nonumber\\
 &=
 \sum_{v\in V(\gamma)} H^{ab}_i (v) \left(1- \hat{h}^i_{\alpha_{ab}} \right) |T_c\rangle,
\end{align}
where 
\begin{equation}\label{expression of H}
H^{ab}_i (v) : = \frac{-9i \hbar^{\frac{1}{2}}}{16\pi^2 \kappa^{\frac{5}{2}} \beta^{\frac{3}{2}}}\; N(v) \epsilon_{ijk} \left(\sum_{e_I \cap \{v\}\ne \emptyset} q^j_I \hat{e}^a_I \right) \left(\sum_{e_J \cap \{v\}\ne \emptyset} q^k_J \hat{e}^b_J \right)  \Upsilon (v).
\end{equation}
When $\epsilon \to 0$, the convergence holds with respect to the so-called URS topology using the diffeomorphism invariant states (for more details consult \cite{LQG}). 
\\

Note that, given the arbitrary nature of the function $N$ in (\ref{H[N] operator}), it can be selectively defined to be non-zero at a single vertex of the graph $\gamma$ while vanishing at all other vertices. As a result, we obtain
\begin{equation}\label{Quantum HC}
\hat{H}_v(N) |T_c\rangle:=  H^{ab}_i(v) \left(1- \hat{h}^i_{\alpha_{ab}} \right)  |T_c\rangle = 0, \;\;\;\;\; \forall v \in V(\gamma),
\end{equation}
as the quantum Hamiltonian constraint.
The states $|T_c\rangle$ satisfying (\ref{Quantum HC}) are called physical states. As is well-known, the Hamiltonian constraint operator is an unbounded, non-symmetric operator and its domain, denoted by $D(\hat{H})$, is dense in $\mathcal{H}_{\text{kin}}$. All states considered henceforth are assumed to lie in $D(\hat{H})$.

It is crucial to emphasize that the quantization procedure employed herein adheres to the standard approach outlined in \cite{Thiemann:1996aw}. Specifically, the quantization of the function (\ref{Classical Hamiltonian Constraint}) is performed via a ``graph-changing'' method, wherein an auxiliary edge is introduced to approximate the curvature at a vertex $v$. This auxiliary edge connects two existing edges emanating from $v$. 
The addition of this edge concurrently introduces two extra vertices to the graph. Specifically, the Hamiltonian constraint appends an edge and two vertices, both of which are trivalent and planar, to the underlying graph.
It is essential to note that this method precludes the utilization of pre-existing graph edges for curvature approximation. Moreover, if the holonomies along the newly added loops $\hat{h}^i_{\alpha_{ab}}$ in (\ref{Quantum HC}) could deduce the valence of the vertex $v$, the adjoint of the Hamiltonian constraint operator would be ill-defined \cite{Assanioussi:2015gka, Yang:2015zda}. This problem can be solved by using the freedom of choosing the representation of  $\hat{h}^i_{\alpha_{ab}}$ to ensure that the valence of $v$ will keep unchanged. This point will be recurrent in the subsequent section.

We introduce an operator $\mathscr{P}$ acting on the Hilbert space $\mathcal{H}_{\text{kin}}$. This operator is defined such that, when applied to a quantum state $T_c$, it eliminates all planar trivalent vertices, henceforth referred to as ``Hamiltonian vertices'', along with any edges connecting these vertices. The set of all states resulting from the application of operator $\mathscr{P}$ to every state in the Hilbert space $\mathcal{H}_{\text{kin}}$ is defined as the ``base Hilbert space'' and denoted by $\mathcal{H}_B$, i.e.
\begin{equation}
\mathcal{H}_B := \mathscr{P} \mathcal{H}_{\text{kin}},
\end{equation}
whose inner product is defined as $\langle T_c, T_{c'}\rangle_B := \delta_{c, c'}$ for all $T_c, T_{c'} \in \mathcal{H}_B$.

\section{\textsf{Shr\"{o}dinger-like equation for the Hamiltonian constraint}}
Our objective in this section is to reformulate (\ref{Quantum HC}) to resemble the Schr\"{o}dinger equation, allowing us to introduce a time parameter into our framework. By adopting the Dirac quantization approach, we have systematically quantized all canonical variables, in contrast to the reduced phase space quantization. This enables us to construct a quantum operator corresponding to the time parameter. Furthermore, we delve into the question of whether this quantum time operator is self-adjoint. To this end, we adopt a strategy analogous to the reduced phase space approach, attempting to solve (\ref{Quantum HC}) for one of the holonomies and then identifying the ``conjugate'' variable to this holonomy as the time parameter. The distinction from the reduced phase space approach lies in the fact that our approach is implemented at the quantum level, whereas the reduced phase space quantization solves the constraint at the classical level.

\subsection{\textsf{Flux representation and discrete dynamics}}
We define the index set $\mathcal{S} = \{(i, a, b) | i\in \{1, 2, 3\}, a, b \in \{x, y, z\}\}$.
Without loss of generality we assume that $H^{xy}_1(v) \ne 0$, then we can rewrite (\ref{Quantum HC}) as

\begin{equation}\label{Isolation of holonomy}
  \left( \hat{h}^1_{\alpha_{xy}} - 1 \right)  |T_c\rangle = - \frac{1}{2 H^{xy}_1(v) } \sum_{\mathcal{S}-\{(1, x,y), (1, y, x)\}} H^{ab}_i(v) \left(1- \hat{h}^i_{\alpha_{ab}} \right)  |T_c\rangle, \;\;\;\;\; \forall v \in V(\gamma).
\end{equation}
We now transition to the flux representation. This representation was introduced in \cite{Baratin:2010nn} through the application of noncommutative group Fourier transform techniques outlined in \cite{Freidel:2005me}. In \cite{Baratin:2010nn}, it is demonstrated that this representation establishes a unitary equivalence for states defined on given graphs (cylindrical functions). Subsequently, the cylindrical consistency of this representation is proved, and hence the continuum limit and the complete LQG Hilbert space can be achieved. Furthermore, the analogous construction in the simpler context of $U(1)$ is also explored in \cite{Baratin:2010nn}, which is relevant to this paper.
We proceed to summarize this Fourier transform in the context of the $U(1)^3$ model. 

Geometrically,  the phase space associated with a single edge is the cotangent bundle $T^* U(1)^3 \simeq U(1)^3 \times \mathfrak{u}(1)^3$. Consequently, for a general graph, the phase space is a product of cotangent bundles $T^* U(1)^3$ over its edges.
The Fourier transform $\mathcal{F}$ maps the Hilbert space $L^2(U(1)^3, \mathrm{d}\mu_H)$ isometrically onto the space $L^2_\star(\mathbb{R}^3, \mathrm{d}\mu)$ of functions on the Lie algebra $\mathfrak{u}(1)^3 \sim \mathbb{R}^3$ endowed with the standard Lebesgue measure $\mathrm{d}\mu$ and a star-product. 
Analogous to the standard Fourier transform on $\mathbb{R}^3$, the construction of $\mathcal{F}$ is rooted in the definition of plane waves
\begin{equation}
\mathsf{e}_g: \mathbb{R}^3 \to U(1), \quad \mathsf{e}_g(\vec{u}) = e^{-i\vec{\phi}_g \cdot \vec{u}},
\end{equation}
where $\vec{\phi}_g = (\phi^1_g, \phi_g^2, \phi_g^3)$ is the coordinate of $g$ on the group manifold $U(1)^3$. For a given choice of coordinates, $\mathcal{F}$ is defined on $L^2(U(1)^3)$ as
\begin{equation}
\mathcal{F}(f)(\vec{u}) = \int \mathrm{d}g f(g) \mathsf{e}_g(\vec{u}),
\end{equation}
where $\mathrm{d}g$ is the normalized Haar measure on the group. 

The Fourier transform can be extended to functions on graphs. Given a graph $\gamma$ with edges $\{e_I\}_{I=1}^{n}$, the corresponding Hilbert space is $L^2((U(1)^3)^n)$, where the measure is the product Haar measure. For $(g_I)_{I=1}^{n} \in (U(1)^3)^n$, the Fourier transform $\mathcal{F}_\gamma$ on the space $L^2((U(1)^3)^n)$ is defined as
\begin{equation}
\mathcal{F}_\gamma (f_\gamma) \big(\mathbf{u}\big) := \int \left(\prod_{I=1}^n \mathrm{d}g_I\right)\; f_\gamma \left((g_I)_{I=1}^n\right) \; \left( \prod_{I=1}^n \mathsf{e}_{g_I} (\vec{u}_I) \right),
\end{equation}
where $\mathbf{u}:=(\vec{u}_I)_{I=1}^{n} \in \left(\mathbb{R}^3\right)^n \sim (\mathfrak{u}(1)^3)^n$. 

As expected, the new representation manifests flux operators as $\star-$multiplications, while holonomies act as translation operators.  In particular, for the holonomy operator $\hat{h}^1_I$ in $q^1_I-$representation and $\psi_\gamma (\mathbf{u}) := \mathcal{F}_\gamma (f_\gamma)(\mathbf{u}) $, the holonomy operator acts dually on $L^2_\star (\mathbb{R}^3)^{\otimes n}$ as $\hat{h}^1_I \psi_\gamma (\mathbf{u}) := \mathcal {F}_\gamma (\hat{h}^1_I f_\gamma)$. If $e_I$ belongs to the underlying graph, we have
\begin{equation}\label{hol op of u 1 in flux rep}
(\hat{h}_I^1 \;\psi_\gamma) (\mathbf{u}) = \psi_\gamma (\vec{u}_1, \cdots, (u_I^1 + q_I^1, u_I^2, u_I^3), \cdots, \vec{u}_n).
\end{equation}
Since (\ref{hol op of u 1 in flux rep}) represents the holonomy as a translation operator, its adjoint acts like
\begin{equation}\label{ad of hol op of u 1 in flux rep}
((\hat{h}_I^1)^\dagger \;\psi_\gamma) (\mathbf{u}) = \psi_\gamma (\vec{u}_1, \cdots, (u_I^1 - q_I^1, u_I^2, u_I^3), \cdots, \vec{u}_n).
\end{equation}
For an edge $e_0$ not belonging to the underlying graph $\gamma$, adding the edge to the graph and applying the holonomy operator $\hat{h}_0^1$ in $q_0^1$-representation results in
\begin{align}
(\hat{h}_0^1 \;\psi_\gamma) (\vec{u}_0, \mathbf{u}) &= \mathcal{F}_{\gamma \cup e_0} (\hat{h}^1_0 f_\gamma) (\vec{u}_0, \mathbf{u}) \nonumber \\
&= \int \left(\mathrm{d}g_0 \prod_{I=1}^n \mathrm{d}g_I\right)\; h^1_0 f_\gamma \left((g_I)_{I=1}^n\right) \; \left(\mathsf{e}_{g_0} (\vec{u}_0) \prod_{I=1}^n \mathsf{e}_{g_I} (\vec{u}_I) \right)\nonumber \\
&= \left(\int  \mathrm{d}g_0\; h^1_0 \; \mathsf{e}_{g_0} (\vec{u}_0) \right)\;  \mathcal{F}_\gamma (f_\gamma) (\mathbf{u}) \nonumber \\
&= \left(\int  \left(\prod_{i=1}^3 \mathrm{d} \phi_0^i \right)\; e^{-i\; {}^1\vec{q}_0 \cdot  \vec{\phi_0}} \; e^{-i\vec{\phi}_0 \cdot \vec{u}_0} \right)\;  \mathcal{F}_\gamma (f_\gamma) (\mathbf{u}) \nonumber \\
&=: \psi_0 (u^1_0+q_0^1, u_0^2, u_0^3) \; \psi_\gamma (\mathbf{u}) \nonumber\\
&=:  \psi_{\gamma \cup e_0} \left((u^1_0+q_0^1, u_0^2, u_0^3), \mathbf{u}\right),
\end{align}
where we have assumed $h^1_0 = g_0 = e^{-i\; {}^1\vec{q}_0 \cdot  \vec{\phi_0}} $ and used the simplified notation $\vec{\phi}_0$ for $\vec{\phi}_{g_0}$. In this case, the adjoint acts like
\begin{equation}\label{ad of hol op of u 1 in flux rep2}
((\hat{h}_0^1)^\dagger \;\psi_\gamma) (\vec{u}_0, \mathbf{u}) =  \psi_{\gamma \cup e_0} \left((u^1_0-q_0^1, u_0^2, u_0^3), \mathbf{u}\right).
\end{equation}
To simplify our analysis, we can assume that all edges added to the graph $\gamma$ due to the Hamiltonian constraint were initially present but labeled with zero. Consequently, instead of modifying $\gamma$, we can start with a complete graph $\gamma' = \gamma \cup \{\alpha_{ab}\}$. 
\\

Let $\psi_{\gamma'} (\mathbf{u}) \in D(\hat{H})\subset \; L^2_\star (\mathbb{R}^3)^{\otimes n}$ be the state Fourier transformed from the charge network state $T_{c'}$, Eq. (\ref{Isolation of holonomy}) in flux representation is
\begin{equation}\label{Isolation of holonomy FR}
 \left( \hat{h}^1_{\alpha_{xy}} - 1 \right)  \psi_{\gamma'} (\mathbf{u}) = \frac{1}{2 H^{xy}_1(v) } \sum_{\mathcal{S}-\{(1, x,y), (1, y, x)\}} H^{ab}_i(v) \left(1- \hat{h}^i_{\alpha_{ab}} \right) \psi_{\gamma'} (\mathbf{u}), \;\;\;\;\; \forall v \in V(\gamma).
\end{equation}
Using the adjoint operators (\ref{ad of hol op of u 1 in flux rep}) and (\ref{ad of hol op of u 1 in flux rep2}), the symmetric Hamiltonian constraint operator which is the symmetric version of (\ref{Isolation of holonomy FR}) can be defined by
\begin{align}\label{self-adjoint Isolation of holonomy FR} 
\frac{1}{2}\left[ \left( \hat{h}^1_{\alpha_{xy}} - 1 \right)\right. &+ \left. \left( \left(\hat{h}^1_{\alpha_{xy}}\right)^\dagger - 1 \right)\right]  \psi_{\gamma'} (\mathbf{u}) \nonumber\\
&= \frac{1}{4 H^{xy}_1(v) } \sum_{\mathcal{S}-\{(1, x,y), (1, y, x)\}} H^{ab}_i(v) \left[ \left(1- \hat{h}^i_{\alpha_{ab}} \right) + \left(1- \left(\hat{h}^i_{\alpha_{ab}}\right)^\dagger \right) \right] \psi_{\gamma'} (\mathbf{u}),
\end{align}
for all  $v \in V(\gamma)$. 
Note that all coeffitients $\frac{H^{ab}_i(v)}{2 H^{xy}_1(v)}$ are real, so the right hand side of (\ref{self-adjoint Isolation of holonomy FR}) is manifestly self-adjoint.
\\
\\
 Now, we are going to calculate $\hat{h}^1_{\alpha_{xy}} \psi_{\gamma'} (\mathbf{u})$. As $\alpha_{xy} = e_x \circ e_{xy} \circ e_y^{-1}$, with the assumption that the first component of the labels of $e_x, e_{xy}$ and $e_y$ are $Q$, we have
 \begin{align}
(\hat{h}^1_{\alpha_{xy}} \psi_{\gamma'} ) (\vec{u}_x, \vec{u}_{xy}, \vec{u}_y, \cdots) &= (\hat{h}^1_{e_x} \cdot  \hat{h}^1_{e_{xy}} \cdot \hat{h}^1_{e_{y}^{-1}} \psi_{\gamma'} ) (\vec{u}_x, \vec{u}_{xy}, \vec{u}_y, \cdots)  \nonumber \\
&= 
\hat{h}^1_{e_x} \psi_{\gamma'}  (\vec{u}_x, (u_{xy}^1 + Q, u_{xy}^2, u_{xy}^3), (u_y^1 - Q, u_y^2, u_y^3), \cdots)  \nonumber \\
&=
 \psi_{\gamma'}  ((u_x^1 + Q, u_x^2, u_x^3), (u_{xy}^1 + Q, u_{xy}^2, u_{xy}^3), (u_y^1 - Q, u_y^2, u_y^3), \cdots).
 \end{align}
By introducing the following variables associated with the loop, the subsequent calculations become more tractable
\begin{equation}\label{new variables}
\vec{w}_1 = \vec{u}_x + \vec{u}_y + \vec{u}_{xy}, \;\;\; \vec{w}_2 = \vec{u}_x- \vec{u}_{xy}, \;\;\; \vec{w}_3 = \vec{u}_y + \vec{u}_{xy}.
\end{equation} 
In fact, using these new variables, it is straightforward to verify that
\begin{align}
(\hat{h}^1_{\alpha_{xy}} \psi_{\gamma'} ) (\vec{w}_1, \vec{w}_2, \vec{w}_3, \cdots) =  \psi_{\gamma'}  ((w_1^1 + Q, w_1^2, w_1^3), \vec{w}_2, \vec{w}_3, \cdots).
\end{align}
By considering (\ref{ad of hol op of u 1 in flux rep}), one can immediately conclude that
\begin{equation}
((\hat{h}^1_{\alpha_{xy}})^\dagger \psi_{\gamma'} ) (\vec{w}_1, \vec{w}_2, \vec{w}_3, \cdots) = \psi_{\gamma'}   ((w_1^1 - Q, w_1^2, w_1^3), \vec{w}_2, \vec{w}_3, \cdots).
\end{equation}
As can be seen, the advantage of employing variables (\ref{new variables}) is that the shift occurs solely in variable $w^1_1$, leaving the remaining variables unaffected.
Note that $Q\in \mathbb{Z}$ is the first component of the label of the newly added edge $e_{xy}$ and is arbitrary. The label assigned to the newly added edge remains somewhat ambiguous in the context of loop quantum gravity which is called spin (charge) representation ambiguity \cite{Perez:2005fn}. However, it is conventionally considered the fundamental representation of the gauge group, provided that that the valence of the vertex $v$ will not be deduced by the action of $\hat{h}_\alpha$ and $\hat{h}_\alpha^\dagger$\footnote{For a recent discussion about the spin repersentation ambiguity, see \cite{Varadarajan:2021zrk}}. Here, we assign a relatively small value of $Q$ to the first component of this label.

By defining 
\begin{align}
t_v &:= w^1_1 = u^1_x + u^1_y + u^1_{xy}\label{dis time}\\
\hat{M}(v) & :=  \frac{1}{4 H^{xy}_1(v) } \sum_{\mathcal{S}-\{(1, x,y), (1, y, x)\}} H^{ab}_i(v) \left[ \left(1- \hat{h}^i_{\alpha_{ab}} \right) + \left(1- \left(\hat{h}^i_{\alpha_{ab}}\right)^\dagger \right) \right],
\end{align}
we can rewrite Eq. (\ref{self-adjoint Isolation of holonomy FR}) as
\begin{equation}\label{Isolation of holonomy FR diff}
\frac{1}{2} \left[ \psi_{\gamma'} \left(t_v+Q, \cdots \right) - \psi_{\gamma'} (t_v, \cdots) +  \psi_{\gamma'} \left(t_v-Q, \cdots \right) - \psi_{\gamma'} (t_v, \cdots) \right] = \hat{M}(v)  \psi_{\gamma'} (t_v, \cdots), \;\;\;\;\; \forall v \in V(\gamma).
\end{equation}

\subsection{\textsf{Shr\"{o}dinger-like equation and its solutions in continuous limit}}
Recall that the operator $\hat{M}(v)$ is self-adjoint. 
We assume that $Q\ne 0$ and then we divide both sides of (\ref{Isolation of holonomy FR diff}) by $Q^2$ to obtain
\begin{equation}\label{Isolation of holonomy FR diff divided by t}
\frac{1}{2}\left[\frac{  \psi_{\gamma'} \left(t_v+Q, \cdots \right) - 2\psi_{\gamma'} (t_v, \cdots) +  \psi_{\gamma'} \left(t_v-Q, \cdots \right)}{Q^2} \right] = \frac{\hat{M}(v)}{Q^2}  \psi_{\gamma'} (t_v, \cdots), \;\;\;\;\; \forall v \in V(\gamma).
\end{equation}
Note that the parameter $Q \in \mathbb{Z}$ is discrete. 
The left-hand side of the preceding equation is reminiscent of a second-order derivative, with the caveat that we can only approach $t_v$ in discrete steps of $Q$. Indeed, upon closer inspection, this is quite natural. As we will demonstrate subsequently, the variable $t_v$ possesses a quantum counterpart whose eigenvalues are precisely $Q \in \mathbb{Z}$. Consequently, the arbitrary variable $Q$ in the left hand side of (\ref{Isolation of holonomy FR diff divided by t}) is restricted to the values that $t_v$ is permitted to assume. Therefore, if we seek to approximate the aforementioned equation in the continuous limit, we obtain
\begin{equation}\label{second derivative Schrodinger equation1}
-\frac{\partial}{\partial t_v} \frac{\partial}{\partial t_v} \psi_{\gamma'} (t_v, \cdots) \approx - \frac{2 \hat{M}(v)}{t_v^2}  \psi_{\gamma'} (t_v, \cdots), \;\;\;\;\; \forall v \in V(\gamma).
\end{equation}
In order to derive an equation involving a first-order derivative operator, we would need to take the square root of both sides of equation (\ref{second derivative Schrodinger equation1}). However, this is only feasible if the operator on the right-hand side is positive. Though this operator is not guaranteed to be positive for all cases, we can restrict our analysis to the subspace of $\mathcal{H}_{\text{kin}}$ corresponding to the negative spectrum of the operator $\hat{M}(v)$. In essence, we are considering only the negative spectral component of the operator. Thus, we obtain
\begin{equation}
-i \frac{\partial}{\partial t_v} \psi_{\gamma'} (t_v, \cdots) \approx \sqrt{- \frac{2 \hat{M}(v)}{t_v^2} } \psi_{\gamma'} (t_v, \cdots), \;\;\;\;\; \forall v \in V(\gamma).
\end{equation}
Therefore, by defining $\hat{\mathbf{H}}(v) := \sqrt{-2 \hat{M}(v)}$, we get
\begin{equation}\label{Schrodinger equation1}
-i \frac{\partial}{\partial t_v} \psi_{\gamma'} (t_v, \cdots) \approx  \frac{\hat{\mathbf{H}}(v)}{|t_v|}  \psi_{\gamma'} (t_v, \cdots), \;\;\;\;\; \forall v \in V(\gamma),
\end{equation}
which resembles the time dependent Schr\"{o}dinger equation.

We aim to determine the parameter $t_v$ in terms of flux variables before solving the equation. In flux representation we expect that the variables $u^i_I$ are in fact the flux variables $E_i (S_{e_I})$ where $S_{e_I}$ is a 2-surface cut by the edge $e_I$. Now based on (\ref{dis time}), we conclude
\begin{equation}\label{time parameter}
t_v = E_1(S_x) + E_1(S_{xy}) + E_1(S_y),
\end{equation}
where $S_x$, $S_{xy}$, and $S_y$ are surfaces that intersect the underlying graph at a single point on edges $e_x$, $e_{xy}$, and $e_y$, respectively, and the orientations are in such a way that $\epsilon(e_{x}, S_x) = \epsilon(e_{xy}, S_{xy})= \epsilon(e_{y}, S_y) = 1$. It is evident that $t_v$ is a gauge-invariant variable. Because the Gauss constraint (as shown in (\ref{Constraints in the model})) is independent of $A$, $E_1 (S)$ is gauge invariant.

We can also justify the form of the time variable (\ref{time parameter}) through an alternative approach. Similar to the reduced phase space quantization, when we solve a constraint for one of the variables and effectively isolate it (as done in (\ref{Isolation of holonomy})), the conjugate variable associated with the isolated variable assumes the role of a time variable. Let us now examine whether the two variables $h^1_{\alpha_{xy}}$ and $t_v$ are conjugate. We begin by calculating the Poisson bracket $\{h^1_{\alpha_{xy}}, t_v \}$:
\begin{align}\label{PB of h and t}
\{h^1_{\alpha_{xy}}, t_v \} &= \{ h^1_{e_x}  h^1_{e_{xy}}  h^1_{e_{y}^{-1}}, t_v \} \nonumber \\
&=
 \{ h^1_{e_x}, E_1 (S_x) \}  h^1_{e_{xy}} h^1_{e_{y}^{-1}} 
 +  h^1_{e_x}  \{h^1_{e_{xy}}, E_1 (S_{xy}) \} h^1_{e_{y}^{-1}}
 +  h^1_{e_x}  h^1_{e_{xy}}  \{h^1_{e_{y}^{-1}}, E_1 (S_y) \} \nonumber \\
 &=
 Q \; h^1_{e_x} h^1_{e_{xy}} h^1_{e_{y}^{-1}} + Q \;  h^1_{e_x} h^1_{e_{xy}} h^1_{e_{y}^{-1}} - Q \; h^1_{e_x} h^1_{e_{xy}} h^1_{e_{y}^{-1}}\nonumber \\
 &=
 Q h^1_{\alpha_{xy}}.
\end{align}
So, $t_v$ is the ``conjugate'' variable associated to  $h^1_{\alpha_{xy}}$ in the sense that $h_e$ is ``conjugate'' to $E(S_e)$ in LQG through the relation $\{h_e, E^i(S_e)\}= \tau^i h_e$.
\\

Let us go back to the Eq. (\ref{Schrodinger equation1}). Note that on the right-hand side of this equation $H^{ab}_i(v)$ exhibits no dependence on the time parameter $t_v$. Upon inspecting the expression for $H^{ab}_i(v)$ in (\ref{expression of H}), one observes the presence of the flux operator (as detailed in (\ref{Flux operator action})) encapsulated within the charges assigned to the graph edges. However, given that the surface $S_{xy}$ interacts exclusively with the newly introduced edge, the operation of $E(S_{xy})$ is contingent solely upon the arbitrary label of the added edge, rendering it impervious to the labels of pre-existing edges, provided that the valence of the vertex $v$ will not be deduced. As a result, the operator $\hat{\mathbf{H}}(v)$ in the right-hand side of (\ref{Schrodinger equation1}) is independent of $t_v$.
It is noteworthy that had we adopted a different operator ordering in (\ref{H[N] operator}), with $\hat{F}$ positioned at rightmost, this result would not have been obtained.

Since the Hamiltonian constraint, responsible for time evolution, acts on the state $|T_c\rangle \in \mathcal{H}_{\text{kin}}$ by adding an edge to the graph, it is intuitively reasonable to assume that the states belonging to the Hilbert space $\mathcal{H}_B$ are those that have not been subjected to this constraint. Let $\psi_0$ be the Fourier transform of a state in $\mathcal{H}_B$, and define a state $|\psi\rangle$ as 
\begin{equation}\label{physical state}
|\psi\rangle_P := \mathrm{e}^{\pm i \ln\left(|t_v|\right) \;\hat{\mathbf{H}}(v)}\; \psi_0,
\end{equation}
where $\psi_0$ is independent of $t_v$, ``$+$'' and ``$-$'' signs correspond to the cases of $t_v >0 $ and $t_v<0$, respectively.
Here, we need to address the well-definedness of the exponential map employed in (\ref{physical state}). If the operator $\hat{\mathbf{H}}(v)$ were bounded, we could simply define the exponential map as $\mathrm{e}^{\hat{\mathbf{H}}(v)} = \sum_{n=0}^\infty \frac{1}{n!} \hat{\mathbf{H}}(v)^n$ and conclude its well-definedness. However, the unbounded nature of $\hat{\mathbf{H}}(v)$ necessitates a more refined treatment to rigorously define Eq. (\ref{physical state}).
For a self-adjoint operator $\hat{A}$ with dense domain $D(\hat{A})$, a dense subspace $D_{\hat{A}} := \bigcap_{n=1}^\infty D(\hat{A}^n) \subseteq D(\hat{A})$ of analytic states exists \cite{Reed:1975uy}. On this subspace, the exponential formula remains valid. However, the series now involves applying the operator to states, i.e.,
\begin{equation}
\mathrm{e}^{\hat{A}}\psi = \sum_{n=0}^{+\infty} \frac{1}{n!}\hat{A}^n \psi\:, \quad \forall \psi \in D_{\hat{A}}.
\end{equation}
One can see that this series is convergent on the said domain \cite{Reed:1975uy}. Therefore, (\ref{physical state}) can be elucidated as
\begin{equation}\label{physical state2}
|\psi\rangle_P := \mathrm{e}^{\pm i \ln\left(|t_v|\right) \;\hat{\mathbf{H}}(v)}\; \psi_0 =  \sum_{n=0}^{+\infty} \frac{(\pm i \ln\left(|t_v|\right))^n}{n!}\hat{\mathbf{H}}(v)^n \psi_0\:, \quad  \psi_0 \in D_{\hat{A}}.
\end{equation}
 It is straightforward to verify that these states satisfy (\ref{Schrodinger equation1}). Consequently, we define the following vector space
\begin{equation}\label{Physical Hilbert Space}
\mathcal{H} = \{|\psi\rangle | |\; \psi\rangle \;\; \text{is a finite linear combination of} \;\; |\psi\rangle_P \}.
\end{equation}
We can equip this vector space with the inner product inherited from $\mathcal{H}_{\text{kin}}$, thus promoting it to a Hilbert space denoted by $(\mathcal{H}, \langle \cdot, \cdot \rangle_{\text{kin}})$. This is the the physical Hilbert space satisfying the Shr\"{o}dinger-like equation (\ref{Schrodinger equation1}) in continuous limit.

\subsection{\textsf{Quantum time operator}}
Given that all canonical variables have corresponding quantum operators within the Dirac quantization framework, we can construct the quantum time operator by directly utilizing the flux operators. With the expression (\ref{time parameter}), we obtain
\begin{equation}\label{quantum time}
\hat{t}_v = \hat{E}_1(S_x) + \hat{E}_1(S_{xy}) + \hat{E}_1(S_y),
\end{equation}
and Eq. (\ref{PB of h and t}) leads to 
\begin{equation}
[\hat{h}^1_{\alpha_{xy}}, \hat{t}_v] = i\hbar \; Q \hat{h}^1_{\alpha_{xy}}.
\end{equation}
A noteworthy aspect of the time operator is its self-adjoint nature, a property inherited from the self-adjoint flux operator. Additionally, it is crucial to note that this operator is instrumental in forming the $\alpha_{xy}$ loops. In essence, the formation of these loops constitutes the ticking of our time clock. As is well-known, when the Hamiltonian constraint acts on a charge network, numerous loops emerge between various edges. However, in our specific case, we have selected the loops formed between links $e_x$ and $e_y$ to serve as our timekeeper, and the evolution of other components of the charge network is measured relative to this time. As previously discussed, the eigenvalues corresponding to the time operator are the values $t_v$, which are restricted to non-zero integers. 
\begin{equation}
\hat{t}_v |T_c \rangle = Q |T_c \rangle.
\end{equation}
So time parameter is discrete in this scenario and we cannot have arbitrary value for time in quantum level.
We can assign the zero time to the states in $\mathcal{H}_B$, which in fact represent the states prior to the imposition of the Hamiltonian constraint.

\section{\textsf{Summary and discussion}}
By closely examining the constraints of the $U(1)^3$ model, one is led to the idea that since all constraints are at most linear in $A$, it may be possible to solve the constraints classically for certain $A$'s and then consider their conjugate variables as time. This is the approach taken in \cite{Thiemann:2022all, Bakhoda:2020ril}. In this paper, we attempt a similar task but at the quantum level. Using Dirac's quantization procedure, we quantize the theory and obtain the quantum operator corresponding to the Hamiltonian constraint. The Hamiltonian operator is observed to be linear in the holonomy operator. By isolating one component of the holonomy and finding its ``conjugate'' variable, the Hamiltonian constraint is expressed as a discrete time evolution of quantum states as shown in Eq. (\ref{Isolation of holonomy FR diff}), as well as a Schr\"{o}dinger-like equation in continuous limit as shown in Eq. (\ref{Schrodinger equation1}). Then the physical sought states that satisfy this Schr\"{o}dinger-like equation are obtained as Eq. (\ref{Physical Hilbert Space}). 

The physical interpretation of our results is rooted in the relational formalism \cite{Rovelli:1990pi}. In this formalism, time evolution of variables is understood as their relative changes in one specific variable. Clearly, the chosen reference variable can be different. Here, we isolate the holonomy $h^1_{\alpha_{xy}}$ and derive the discrete time evolution of quantum states and the Schr\"{o}dinger-like equation based on it in continuous limit, but we could have used any other component of the holonomy. The loop formed between two specific edges of the underlying charge network plays a crucial role, and in fact, time evolution of other variables are measured relative to the quantum number in the added edge to form the loop or the fluxes over the surfaces $S_\alpha$ intersecting it.

It should be noted that, while the Hamiltonian constraint operator in (\ref{Quantum HC}) is defined in the kinematical Hilbert space $\mathcal{H}_{\text{kin}}$, it is straightforward to define it in certain almost diffeomorphism invariant Hilbert space $\mathcal{H}_{np3}$, which consists of those distributional states that are invariant under the diffeomorphism transformations preserving the non-planar vertices with valence higher than 2 of any underlying charge network states, mimicking the construction in  \cite{Yang:2015zda}. This is because the Hamiltonian constraint operator acts only on the non-planar vertices with valence higher than 2 and its regulator can be naturally removed in $\mathcal{H}_{np3}$. Thus, the concept of discrete relative time evolution and the Schr\"{o}dinger-like equation in continuous limit can also be extended to $\mathcal{H}_{np3}$. It is worth further investigating whether an analysis similar to what we did in this paper could be performed in the full theory of LQG. Clearly, the situation will be more complex, as the form of the constraints is more complicated and the gauge group is non-Abelian. Consequently, many of the calculations performed here cannot be directly generalized to the non-Abelian case. We leave this task for further investigation.

\subsection*{\textsf{Acknowledgement}} 
The present work is funded by the National Natural Science Foundation of China (NSFC) under Grants no. 12275022 and no. 11875006.

\appendix

}
\end{document}